\documentclass[%
reprint,
 amsmath,amssymb,
aps,
prb,
]{revtex4-1}

\usepackage{graphicx}
\usepackage{dcolumn}
\usepackage{bm}

\begin{document}

\preprint{APS/123-QED}

\title{Dynamical quantum phase transition in quantum spin chains with gapless phases}

\author{Kaiyuan Cao}
 \affiliation{Department of Physics and Institute of Theoretical Physics, Nanjing Normal University, Nanjing 210023, P. R. China}

\author{Zhong Ming}%
\email{mzhong@njnu.edu.cn}
\affiliation{Department of Physics and Institute of Theoretical Physics, Nanjing Normal University, Nanjing 210023, P. R. China}

\author{Peiqing Tong}
\email{pqtong@njnu.edu.cn}
\affiliation{Department of Physics and Institute of Theoretical Physics, Nanjing Normal University, Nanjing 210023, P. R. China and Jiangsu Key Laboratory for Numerical Simulation of Large Scale Complex Systems, Nanjing Normal University, Nanjing 210023, P. R. China}%

\date{\today}

\begin{abstract}
  The dynamical quantum phase transitions (DQPTs) in quantum spin chains with gapless phases after a sudden quench are studied. We mainly consider the general systems with asymmetrical quasiparticle excitation spectra and obtain the general expression of the Loschmidt echo as well as the general conditions for the occurrence of DQPTs. As two examples, we study the DQPTs in the \emph{XY} chains with Dzyaloshinskii-Moriya interaction and \emph{XZY-YZX} type of three-site interaction. It's found that the DQPTs may not occur in the quench across the quantum phase transitions regardless of whether the quench is from the gapless phase to gapped phase or from the gapped phase to gapless phase. This is different from the DQPTs in the case of quench from the gapped phase to gapped phase, in which the DQPTs will always appear. Besides, we also analyze the different reasons for the absence of DQPTs in the quench from the gapless phase and the gapped phase.
\end{abstract}

\maketitle

%

\section{Introduction}

Quantum phase transitions (QPTs) are one of the most significant phenomena in quantum many-body physics\cite{Vojta2002,Vojta_2003,Sachdev_2000}. It is characterized by a nonanalytic behavior of some physical observable at a quantum critical point due to the change of the external control parameter. Recently, the dynamical quantum phase transitions (DQPTs) that emerge in the dynamics of an isolated quantum many-body systems have attracted both the theoretical\cite{Heyl2013,Karrasch2013,Andraschko2014,Heyl2014,Hickey2014,Kriel2014,Vajna2014,Heyl2015,
Schmitt2015,Vajna2015,Budich2016,Divakaran2016,Huang2016,Sharma2016,Zvyagin2016,Bhattacharya2017,
Halimeh2017,Heyl2017,Heyl2017_1,Homrighausen2017,Karrasch2017,Weidinger2017,Bhattacharjee2018,
Cheraghi_2018,Heyl_2018,Kennes2018,Kosior2018,Kosior2018,Lang2018,Wang2018,Sedlmayr2018,Yin2018,
Zhou2018,Bojan2018,Abdi2019,Hagymasi2019,Huang2019,Jafari2019,Khatun2019,Lacki2019,Lahiri2019,Liu2019,
Mendl2019,Piccitto2019,Srivastav2019,Yang2019,Cao2020,Hou2020,Masowski2020,Haldar2020,Wu2020,Zamani2020,
Jafari2021} and experimental\cite{Jurcevic2017,Zhang2017,Flaschner2018,Wang2019,Nie2020,Tian2020,
Xu2020} interests. Unlike the QPTs driven by the external control parameters, the DQPTs describe the nonanalytic behavior of the Loschmidt echo (LE) during the time evolution, where a common protocol for driven a system out of equilibrium is called quantum quench. In many cases, the DQPTs are found to have a strong connection with the QPTs. The DQPTs are present in the cases of sudden quenches crossing the quantum critical points\cite{Heyl2013,Karrasch2013,Andraschko2014} and the topology changes\cite{Hickey2014,Vajna2015}, although it is also found that the DQPTs occur in the cases of quenches without crossing any quantum critical point\cite{Vajna2014,Halimeh2017,Homrighausen2017}. Therefore, it is still an open and debated issue whether the quench crossing the critical points of QPTs is a necessary condition to induce a DQPT\cite{Jafari2019,Haldar2020}.

Up to now, so many works have investigated the quench from gapped phase to gapped phase, but few studies have focused on the case of quench from the gapless phase. The gapless phase is of general existence in the quantum systems, such as the \emph{XY} chain with the Dzyaloshinskii-Moriya (DM) interaction\cite{Dzyaloshinsky1958,Moriya1960,Derzhko1994,Derzhko1999}, the \emph{XY} chain with \emph{XZY-YZX} type of three-site interactions\cite{Gottlieb1999,Krokhmalskii2008} and Kitaev model\cite{Kitaev2006} etc. Unlike the case of the gapped phase, the ground states of the system in the gapless phase corresponds to the configuration where all the states with negative excitation spectra are filled and nonnegative are empty\cite{Rodney2013,Zhong_2013}. Recently, Cheraghi and Mahdavifar studied the DQPTs in the quantum Ising chain with DM interaction \cite{Cheraghi_2018}. They only considered the case of quench from gapped phase to gapped phase and did not consider the quench from the gapless phase, so that they concluded that the DM interaction does not affect the DQPT. In a recent paper \cite{Haldar2020}, the authors studied the DQPTs in the \emph{XY} chain with DM interaction in the alternating external transverse field. They also did not discuss the case of quench starting from the gapless phase, because they said that it's difficult to study the initial condition in the gapless phase in their models \cite{Haldar2020}. Therefore, it's still unknown when the quench is from the gapless phase.

Recently, Jafari has studied the DQPTs in the extended \emph{XY} chain with \emph{XZX+YZY} type of three-site interaction in the staggered transverse field \cite{Jafari2019}. He found that the DQPTs may not occur when the quench is from the gapped phase to gapless phase. The model has the gapless phase due to the staggered external field and the quasiparticle excitation spectra are symmetrical \cite{Jafari2019}. In this paper, we investigate the DQPTs in more general quantum spin chains with  asymmetrical quasiparticle excitation spectra. We study all possible situations of ground states of pre-quench Hamiltonian and obtain the general expression of LE for the systems with gapless phases [see Sec.~\uppercase\expandafter{\romannumeral2}]. For the homogeneous system, the LE can be given by $\mathcal{L}(t)=\prod_{k>0}\mathcal{L}_{k}(t)$, where $\mathcal{L}_{k}(t)$ equal unity corresponding to the quasiparticle excitation spectra of pre-quench Hamiltonian satisfying $\varepsilon_{k}\cdot\varepsilon_{-k}<0$ in the momentum subspace $k>0$. The general conditions for the occurrence of DQPTs are also obtained in Sec.~\uppercase\expandafter{\romannumeral2}. In Secs.~\uppercase\expandafter{\romannumeral3} and \uppercase\expandafter{\romannumeral4}, we discuss the behaviors of DQPTs in two typical models---the \emph{XY} chains with DM and \emph{XZY-YZX} type of three-site interactions, respectively.  It's found that the DQPTs may not occur in the quench across the quantum phase transitions regardless of whether the quench is from the gapless phase to gapped phase or from gapped phase to gapless phase. This is different from that in the quantum spin chain with symmetrical excitation spectra.  A brief conclusion is given in Sec.~\uppercase\expandafter{\romannumeral5}.

\section{Models and DQPT}

We consider general quantum spin chains, whose Hamiltonian can be written in a quadratic form
\begin{equation}\label{quardratic_H}
  H=\sum_{nm}c^{\dag}_{n}A_{nm}c_{m}+\frac{1}{2}\sum_{nm}(c^{\dag}_{n}B_{nm}c^{\dag}_{m}+h.c.),
\end{equation}
where $c_{n}$ and $c^{\dag}_{n}$ are fermion annihilation and creation operators respectively. The Hermiticity of $H$ demands $A$ to be a Hermitian matrix and anti-commutation of fermion operators demands $B$ to be an antisymmetric matrix. In the uniform case, the Hamiltonian (\ref{quardratic_H}) can be written in a diagonal form in the momentum space via the Fourier transformation ($c_{k}=\frac{1}{\sqrt{N}}\sum_{n}e^{ikn}c_{n}$, $c^{\dag}_{k}=\frac{1}{\sqrt{N}}\sum_{n}e^{-ikn}c^{\dag}_{n}$) and Bogoliubov transformation \cite{Sachdev_2000,Suzuki2013} $\eta_{k}=u_{k}c_{k}+iv_{k}c^{\dag}_{-k}$:
\begin{equation}\label{diagonal_H}
  \begin{split}
    H & =\sum_{k}\varepsilon_{k}(\eta^{\dag}_{k}\eta_{k}-\frac{1}{2}) \\
    & =\sum_{k>0}[\varepsilon_{k}(\eta^{
    \dag}_{k}\eta_{k}-\frac{1}{2})+\varepsilon_{-k}(\eta^{\dag}_{-k}\eta_{-k}-\frac{1}{2}] \\ & =\sum_{k>0}H_{k},
  \end{split}
\end{equation}
where $k$ are the waves vectors, $\varepsilon_{k}$ are the quasiparticle excitation spectra, $\eta_{k}$ and $\eta^{\dag}_{k}$ are fermion annihilation and creation operators. It should be noticed that we consider the more general cases in which the quasiparticle excitation spectra may not be asymmetrical ($\varepsilon_{k}\neq\varepsilon_{-k}$) and positive ($\varepsilon_{k}<0$).

The ground state of the system can be constructed as
\begin{equation}\label{ground_state}
  |GS\rangle=\bigotimes_{k>0}|GS\rangle_{k}.
\end{equation}
Here $|GS\rangle_{k}\in\mathbb{C}^{2}\bigotimes\mathbb{C}^{2}$ in the subspace $k>0$ is of the form
\begin{equation}\label{two_site_state}
  |GS\rangle_{k}=|n_{k}n_{-k}\rangle,
\end{equation}
where $n_{k}(n_{-k})=0$ or $1$ are the eigenvalues of spinless fermion number operator $\hat{n}_{k}=\eta^{\dag}_{k}\eta_{k}(\hat{n}_{-k}=\eta^{\dag}_{-k}\eta_{-k})$. To be specific, the ground states are given by
\begin{equation}\label{ground_states}
   |GS\rangle_{k}=\left\{\begin{array}{cl}
                          |0_{k}0_{-k}\rangle,  & \varepsilon_{k},\varepsilon_{-k}\geq0 \\
                          |0_{k}1_{-k}\rangle, & \varepsilon_{k}\geq0,\varepsilon_{-k}<0 \\
                          |1_{k}0_{-k}\rangle, & \varepsilon_{k}<0,\varepsilon_{-k}\geq0 \\
                          |1_{k}1_{-k}\rangle, & \varepsilon_{k},\varepsilon_{-k}<0.
                        \end{array}
  \right.
\end{equation}

In a quantum quench, the system is prepared in the ground states $|GS\rangle=\otimes_{k>0}|n_{k}n_{-k}\rangle$ for the Hamiltonian $H$. At time $t=0$, the system undergoes a sudden quench with its Hamiltonian switched from $H$ to $\widetilde{H}$, where the ground states of the post-quench Hamiltonian $\widetilde{H}$ is $|\widetilde{GS}\rangle=\otimes_{k>0}|\widetilde{n}_{k}\widetilde{n}_{-k}\rangle$ accordingly. The quantities associated with the post-Hamiltonian are labeled by sign ``$\sim$''. The Loschmidt amplitude is defined as (we take $\hbar=1$)
\begin{equation}\label{Loschmidt_amplitude}
    {\cal G}(t) =\langle GS|e^{iHt}e^{-i\widetilde{H}t}|GS\rangle  =e^{-i(E_{0}-\widetilde{E}_{0})t}\prod_{k>0}{\cal G}_{k}(t),
\end{equation}
where ${\cal G}_{k}(t)=\langle n_{k}n_{-k}|e^{-i\widetilde{H}_{k}t}|n_{k}n_{-k}\rangle$, $E_{0}(\widetilde{E}_{0})$ is the ground state energy.
To calculate the Loschmidt amplitude, we should expand the states $|n_{k}n_{-k}\rangle$ of the pre-quench Hamiltonian $H_{k}$ by the eigenstates of the post-quench $\widetilde{H}_{k}$.

By considering both $\eta^{\dag}_{k}$ and $\widetilde{\eta}^{\dag}_{k}$ are related by the Bogoliubov transformation, we can obtain the relations between the eigenstates of $H_{k}$ and $\widetilde{H}_{k}$:
\begin{equation}\label{eigenstates}
  \begin{array}{l}
    |0_{k}0_{-k}\rangle=\cos{\alpha_{k}}|\widetilde{0}_{k}\widetilde{0}_{-k}
    \rangle-i\sin{\alpha_{k}}|\widetilde{1}_{k}\widetilde{1}_{-k}\rangle \\
    |1_{k}0_{-k}\rangle=|\widetilde{1}_{k}\widetilde{0}_{-k}\rangle \\
    |0_{k}1_{-k}\rangle=|\widetilde{0}_{k}\widetilde{1}_{-k}\rangle \\
    |1_{k}1_{-k}\rangle=-i\sin{\alpha_{k}}|\widetilde{0}_{k}\widetilde{0}_{-k}
    \rangle+ \cos{\alpha_{k}}|\widetilde{1}_{k}\widetilde{1}_{-k}\rangle,
  \end{array}
\end{equation}
where $\alpha_{k}=\theta_{k}-\tilde{\theta}_{k}$, $\theta_{k}$ ($\tilde{\theta}_{k}$) are the Bogoliubov angles of the pre- and post-quench Hamiltonian defined by $u_{k}=\cos{\theta_{k}}$ and $v_{k}=\sin{\theta_{k}}$ \cite{Sachdev_2000,Suzuki2013}. It's noticed that $|1_{k}0_{-k}\rangle$ and $|0_{k}1_{-k}\rangle$ of the pre-quench Hamiltonian are also the eigenstates of the post-quench Hamiltonian $\widetilde{H}_{k}$.

By substituting Eq.~(\ref{eigenstates}) into Eq.~(\ref{Loschmidt_amplitude}), the Loschmidt amplitudes ${\cal G}_{k}(t)$  are given by
\begin{equation}\label{LA_k}
  {\cal G}_{k}(t)=\left\{\begin{array}{cr}
                           \cos^{2}{\alpha_{k}}+\sin^{2}{\alpha_{k}}e^{-it(
                           \widetilde{\varepsilon}_{k}+\widetilde{\varepsilon}_{-k})}, & \varepsilon_{k},\varepsilon_{-k}\geq0 \\
                           e^{-it(
                           \widetilde{\varepsilon}_{k}+\widetilde{\varepsilon}_{-k})}, & \varepsilon_{k}<0,\varepsilon_{-k}\geq0 \\
                           e^{-it(
                           \widetilde{\varepsilon}_{k}+\widetilde{\varepsilon}_{-k})}, & \varepsilon_{k}\geq0,\varepsilon_{-k}<0 \\
                           \sin^{2}{\alpha_{k}}+\cos^{2}{\alpha_
  {k}}e^{-it(\widetilde{\varepsilon}_{k}+\widetilde{\varepsilon}_{-k})}, & \varepsilon_{k},\varepsilon_{-k}<0
                         \end{array}
  \right.
\end{equation}
Then, the LE is
\begin{equation}\label{Loschmidt_echo}
    {\cal L}(t) =|{\cal G}(t)|^{2} =\prod_{k>0}{\cal L}_{k}(t)
\end{equation}
with
\begin{equation}\label{LO_k}
  {\cal L}_{k}(t)=\left\{\begin{array}{cr}
                           1-\sin^{2}{2\alpha_{k}}\sin^{2}{\frac{
                           \widetilde{\varepsilon}_{k}+\widetilde{\varepsilon}_{-k}}{2}t}, & \varepsilon_{k}\cdot\varepsilon_{-k}\geq0, \\
                           1, & \varepsilon_{k}\cdot\varepsilon_{-k}\leq0.
                         \end{array}
  \right.
\end{equation}

In order to show the DQPT more directly, one can use the dynamical free energy density, which is defined as the rate function $\lambda(t)=-\lim_{N\rightarrow\infty}\ln{{\cal L}(t)}/N$ of the LE. Analogous to the equilibrium phase transitions, the rate function will show the sharp nonanalyticities at the critical times of the DQPTs \cite{Karrasch2013}. According to Eq.~(\ref{LO_k}), the rate function of LE is
\begin{equation}\label{rate_function}
  \lambda(t)=-\lim_{N\rightarrow\infty}\frac{1}{N}\sum_{k>0}\ln{{\cal L}_{k}(t)}.
\end{equation}
Obviously, for the cases of $\varepsilon_{k}\cdot\varepsilon_{-k}\leq0$, $\ln{{\cal L}_{k}(t)}=0$.

Another way to show the DQPTs is using the Fisher zeros of the Loschmidt amplitude in the complex time plane \cite{Heyl2013, Heyl_2018}. According to Eq.~(\ref{LA_k}), for $\varepsilon_{k}<0,\varepsilon_{-k}\geq0$ and $\varepsilon_{k}\geq0,\varepsilon_{-k}<0$, ${\cal G}_{k}(t)=e^{-it(\widetilde{\varepsilon}_{k}+\widetilde{\varepsilon}_{-k})}=0$ have no solutions in the complex time plane. While for $\varepsilon_{k},\varepsilon_{-k}\geq0$, we obtain the Fisher zeros $z_{n}$ of the Loschmidt amplitude located on the lines ($n\in\mathbb{Z}$)
\begin{equation}\label{Fisher_1}
  z_{n}(k)=\frac{1}{\widetilde{\varepsilon}_{k}+\widetilde{\varepsilon}_{-k}}
  [\ln{\tan^{2}{\alpha_{k}}}+i\pi(2n+1)],
\end{equation}
and for $\varepsilon_{k},\varepsilon_{-k}<0$
\begin{equation}\label{Fisher_2}
  z_{n}(k)=\frac{1}{\widetilde{\varepsilon}_{k}+\widetilde{\varepsilon}_{-k}}
  [-\ln{\tan^{2}{\alpha_{k}}}+i\pi(2n+1)],
\end{equation}
The DQPTs occur if the Fisher zeros lines interacting with the imaginary axis. The imaginary parts of the Fisher zeros on the imaginary axis denote the critical times of the DQPTs. Therefore, the conditions for the occurrence of DQPTs are given by

\begin{equation}\label{occur_condition}
  \tan^{2}{\alpha_{k}}=1 \quad \text{and} \quad \varepsilon_{k}\cdot\varepsilon_{-k}\geq0
\end{equation}

If the above conditions (\ref{occur_condition}) hold, the corresponding critical times of the DQPTs are given by
\begin{equation}\label{critical_time}
  t_{n}(k)=\frac{\pi}{\widetilde{\varepsilon}_{k}+\widetilde{\varepsilon}
  _{-k}}(2n+1).
\end{equation}

\section{the XY chain with DM interaction}

As the first example, we begin our study on the anisotropic \emph{XY} chain with the DM interaction in the transverse field \cite{Rodney2013,Zhong_2013}
\begin{equation}\label{DM}
 \begin{split}
    H & =-\sum_{n=1}^{N}(\frac{1+\gamma}{2}\sigma^{x}_{n}\sigma^{x}_{n+1}+\frac{1-\gamma}{2}\sigma^{y}_{n}\sigma^{y}
    _{n+1}+h\sigma^{z}_{n}) \\
      & \quad -\sum_{n=1}^{N}D(\sigma^{x}_{n}\sigma^{y}_{n+1}-\sigma^{y}_{n}\sigma^{x}_{n+1}),
  \end{split}
\end{equation}
where $\sigma^{x,y,z}_{n}$ are the Pauli matrices, $\gamma$ is the anisotropic parameter, $h$ is the external field, and $D$ is the strength of the DM interaction, respectively.

The Hamiltonian (\ref{DM}) can be expressed in the quadratic form (\ref{quardratic_H}) via the Jordan-Wigner transformation, written by
\begin{equation}\label{DM_AB}
  \begin{array}{l}
    \begin{aligned}
      A_{mn}= & -h\delta_{mn}-(1/2+iD)\delta_{m,n+1} \\
             &  -(1/2-iD)\delta_{m+1,n},
    \end{aligned}
     \\
    B_{mn}= \gamma\delta_{m,n+1}/2 - \gamma\delta_{m+1,n}/2.
  \end{array}
\end{equation}
The quasiparticle excitation spectra $\varepsilon_{k}$ of the XY chain with DM interaction are given by
\begin{equation}\label{DM_anisotropic_varepsilon}
  \varepsilon_{k}=-4D\sin{k}+2\sqrt{(h+\cos{k})^{2}+\gamma^{2}\sin^{2}{k}}.
\end{equation}
According to Eq.~(\ref{DM_anisotropic_varepsilon}), the quasiparticle excitation spectra $\varepsilon_{k}$ are not symmetrical in the momentum space and can only have negative values for $k>0$.

\begin{figure}[t]
  \centering
  \includegraphics[width=1.0\linewidth]{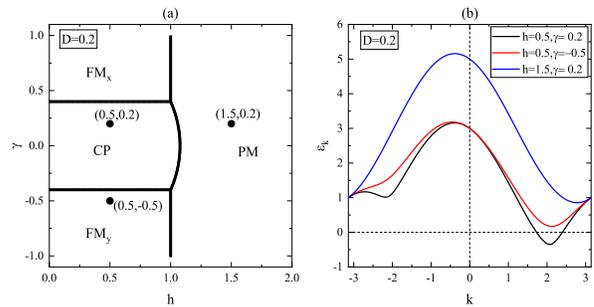}\\
  \caption{(a)The phase diagram of the XY chain with DM interaction for $D=0.2$. Regions CP, PM, FM$_{x}$ and FM$_{y}$ between the solid lines correspond to the gapless chiral phase, gapped paramagnetic phase, gapped ferromagnetic phase along $x$ direction and gapped ferromagnetic phase along $y$ direction, respectively. (b) The typical quasiparticle excitation spectra corresponding to the Hamiltonian parameters marked by black dots in (a).  } \label{DM_phase}
\end{figure}

Fig.~\ref{DM_phase} (a) shows a typical phase diagram of the system with $D=0.2$ in the plane $(h,\gamma)$ \cite{Zhong_2013}, where the region CP is the gapless chiral phase, region PM is the gapped paramagnetic phase, region FM$_{x}$ is the gapped ferromagnetic phase along $x$ direction, FM$_{y}$ is the gapped ferromagnetic phase along $y$ direction, respectively. The critical line separating the gapless chiral phase and the PM phase is given by
$h_{c}=\sqrt{4D^{2}-\gamma^{2}+1}$, and those separating the gapless chiral phase and the FM phases are $\gamma_{c}=\pm2D$. In Fig.~\ref{DM_phase} (b), we show three examples of quasiparticle excitation spectra, in which the Hamiltonian parameters $(h=0.5,\gamma=0.2)$, $(h=0.5,\gamma=-0.5)$ and $(h=1.5,\gamma=0.2)$ are marked by black dots in Fig.~\ref{DM_phase} (a). The excitation spectra $\varepsilon_{k}$ are negative for $1.7<k<2.4$ when $h=0.5,\gamma=0.2$, while they are positive for all $k$ when $h=0.5,\gamma=-0.5$ and $h=1.5,\gamma=0.2$.

\begin{figure}[b]
  \centering
  \includegraphics[width=1.0\linewidth]{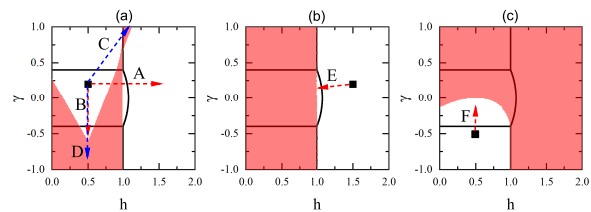}\\
  \caption{The domains ${\cal R}(h_{0},\gamma_{0})$ cover the post-quench Hamiltonian parameters where the DQPTs appear in the case of quench from the initial points $(h_{0},\gamma_{0})$. The initial points are marked by black squared dots: (a) $(h_{0}=0.5,\gamma_{0}=0.2)$, (b) $(h_{0}=1.5,\gamma_{0}=0.2)$, and (c) $(h_{0}=0.5,\gamma_{0}=-0.5)$.}\label{domain_DM}
\end{figure}

Now, we consider the quench from $H=H(h_{0},\gamma_{0})$ to $\widetilde{H}=H(h_{1},\gamma_{1})$, where we keep the same strength of DM interaction before and after the quench. By considering the Bogoliubov angle satisfying $\tan{2\theta_{k}}=\gamma\sin{k}/(h+\cos{k})$, the condition $|\tan{\alpha_{k}}|=1$ of (\ref{occur_condition}) is equivalent to the following quadratic equation
\begin{equation}\label{quard_equation}
  (1-\gamma_{0}\gamma_{1})\cos^{2}{k}+(h_{0}+h_{1})\cos{k}+(h_{0}h_{1}+\gamma_
  {0}\gamma_{1})=0.
\end{equation}
Therefore, for a given quench process, the conditions for the occurrence of DQPTs are given by
\begin{equation}\label{DM_exist_condition}
  \left\{\begin{array}{l}
           \cos{k}=\frac{-(h_{0}+h_{1})\pm\sqrt{\Delta}}{2(1-\gamma_{0}\gamma_
           {1})},  \\
           \varepsilon_{k}\cdot\varepsilon_{-k}\geq0,
         \end{array}
  \right.
\end{equation}
where $\Delta=(h_{0}+h_{1})^{2}-4(1-\gamma_{0}\gamma_{1})(h_{0}h_{1}+\gamma_{0}\gamma_{1})$.

In Fig.~\ref{domain_DM}, we use the domains ${\cal R}(h_{0},\gamma_{0})$ to cover all the post-quench Hamiltonian parameters where the Eqs.~(\ref{DM_exist_condition}) are satisfied in the case of quench from the pre-quench Hamiltonian $H(h_{0},\gamma_{0})$. In other words, the DQPTs occur in the quench from $H(h_{0},\gamma_{0})$ to the post-quench Hamiltonian parameters covered by the domain ${\cal R}(h_{0},\gamma_{0})$.

\begin{figure}[t]
  \centering
  \includegraphics[width=1.0\linewidth]{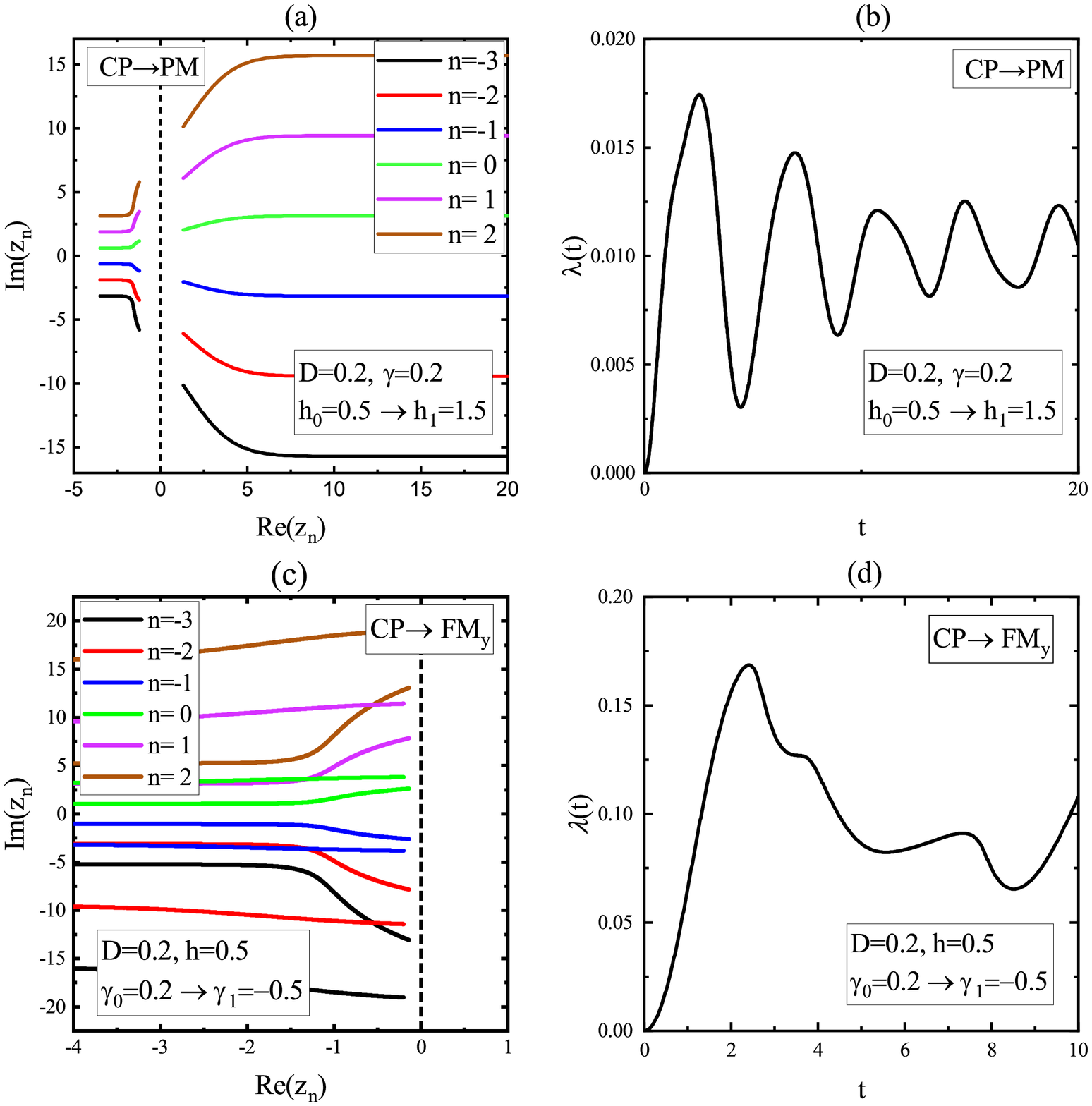}\\
  \caption{The Fisher zeros and corresponding rate functions for the quenches from the gapless phase to the gapped phase in the anisotropic XY chain with DM interaction. The quench in (a) and (b) is from the gapless chiral phase to the PM phase (CP$\rightarrow$PM), that is the path \textbf{A} seen in Fig.~\ref{domain_DM} (a). The quench in (c) and (d) is from the gapless chiral phase to the FM$_{y}$ phase (CP$\rightarrow$FM$_y$), that is the path \textbf{B} seen in Fig.~\ref{domain_DM} (a).}
  \label{DM_anisotropic_gapless}
\end{figure}

\begin{figure}
  \centering
  \includegraphics[width=1.0\linewidth]{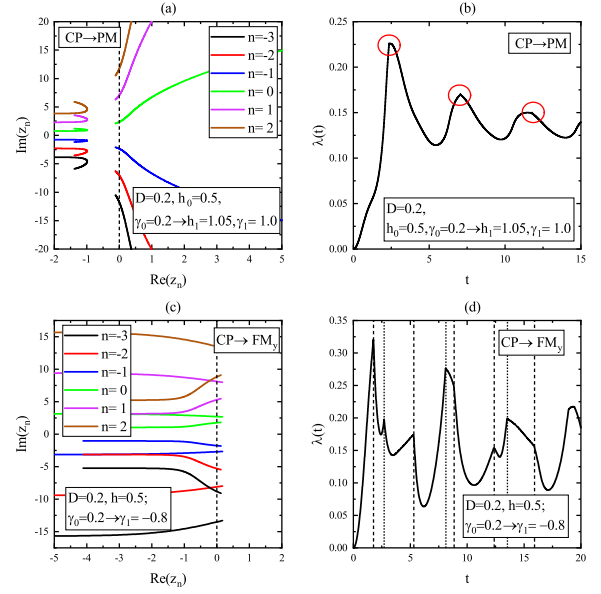}\\
  \caption{The Fisher zeros and corresponding rate functions for the quenches from the gapless phase to the gapped phase in the anisotropic XY chain with DM interaction. The quench in (a) and (b) is from the gapless chiral phase to the PM phase (CP$\rightarrow$PM), that is the path \textbf{C} seen in Fig.~\ref{domain_DM} (a). The quench in (c) and (d) is from the gapless chiral phase to the FM$_{y}$ phase (CP$\rightarrow$FM$_y$), that is the path \textbf{D} seen in Fig.~\ref{domain_DM} (a). Two different groups of critical times are labeled by the short-dash lines and short dot lines respectively in (d). }\label{DM_anisotropic_gapless1}
\end{figure}

\paragraph*{Quench from the gapless phase to gapped phase.} In Fig.~\ref{domain_DM} (a), we display a typical example for the quench from the pre-quench Hamiltonian $H(h_{0}=0.5,\gamma_{0}=0.2)$. The domain $\mathcal{R}(0.5,0.2)$ doesn't cover all the regions of PM, FM$_x$, and FM$_{y}$ phases. This indicates that
it is possible that the DQPTs do not occur when the quench crosses the critical points from gapless phase to gapped phase.

To show the behaviors of DQPTs more clearly, we show two examples of the Fisher zeros and the corresponding rate functions for the quenches from gapless phase to gapped phase in Fig.~\ref{DM_anisotropic_gapless}. The results in Figs.~\ref{DM_anisotropic_gapless} (a) and (b) are in the case of quench from $(h_{0}=0.5,\gamma_{0}=0.2)$ to $(h_{1}=1.5,\gamma_{1}=0.2)$ [see the path A in Fig.~\ref{domain_DM} (a)]. It is found that the Fisher zeros lines $z_{n}(k)$ are separated into two parts on both sides of the imaginary axis and have no intersection with the imaginary axis. The reason is that those wave vectors corresponding to $\varepsilon_{k}\cdot\varepsilon_{-k}<0$ do not contribute to the Fisher zeros. Meanwhile, the rate function of the LE shows smooth evolution with time. In another case of quench that is from $(h_{0}=0.5,\gamma_{0}=0.2)$ to $(h_{1}=0.5,\gamma_{1}=-0.5)$ [see path B in Fig.~\ref{domain_DM} (b)], the Fisher zeros lines $z_{n}(k)$ are found separated into two parts on the left of the imaginary axis [see Fig.~\ref{DM_anisotropic_gapless} (c)]. Similarly, there is no singular point in the rate functions.

For comparison, we also show two examples where the DQPTs occur in the cases of quenches from the gapless phase to gapped phase in Fig.~\ref{DM_anisotropic_gapless1}. The results in Figs.~\ref{DM_anisotropic_gapless1} (a) and (b) correspond to the quench from $(h_{0}=0.5,\gamma_{0}=0.2)$ to $(h_{1}=1.05,\gamma_{1}=1.0)$ [see the path C in Fig.~\ref{domain_DM} (a)]. It is found that although the lines of Fisher zeros $z_{n}$ are separated into two parts, the right branches of Fisher zeros interact with the imaginary axis in the complex time plane. Correspondingly, we can see the singular points in the rate function of LE [see Fig.~\ref{DM_anisotropic_gapless1} (b)]. The results in Figs.~\ref{DM_anisotropic_gapless1} (c) and (d) correspond to the quench from $(h_{0}=0.5,\gamma_{0}=0.2)$ to $(h_{1}=0.5,\gamma_{1}=-0.8)$ [see the path D in Fig.~\ref{domain_DM} (a)]. It can be seen that two separated branches of Fisher zeros lines $z_{n}$ both have interactions with the imaginary axis, and two corresponding groups of critical times are marked by the short dash and short dot lines in the rate function [see Fig.~\ref{DM_anisotropic_gapless1} (d)].

Apparently, in the case of quench from the gapless phase, the lines of Fisher zeros will be cut into two separated pieces because the wave vectors corresponding to $\varepsilon_{k}\cdot\varepsilon_{-k}<0$ do not contribute to the Fisher zeros. The occurrence of DQPT requires that the Bogoliubov angles satisfying $|\tan{\alpha_{k}}|=1$ are located outside the wave vector intervals corresponding to $\varepsilon_{k}\cdot\varepsilon_{-k}<0$.

\begin{figure}[t]
  \centering
  \includegraphics[width=1.0\linewidth]{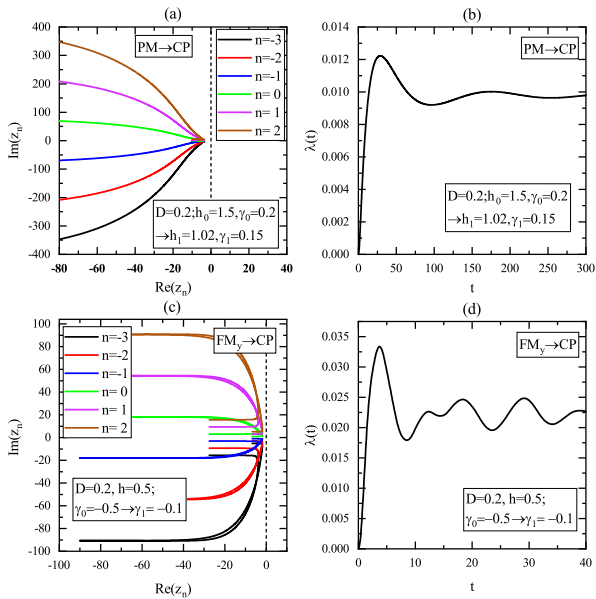}\\
  \caption{The Fisher zeros and corresponding rate functions for the quenches from the gapped phase in the anisotropic XY chain with DM interaction. The quench process in (a) and (b) is from the gapped PM phase to the gapless chiral phase (PM$\rightarrow$CP), that is the path \textbf{E} seen in Fig.~\ref{domain_DM} (b). The quench process in (c) and (d) is from the gapped FM$_{y}$ phase to the gapless chiral phase (FM$_x$$\rightarrow$CP), that is the path \textbf{F} seen in Fig.~\ref{domain_DM} (c).}\label{DM_anisotropic_gapped}
\end{figure}

\paragraph*{Quench from the gapped phase to gapless phase.} Now, we study the case of quench from the gapped phase. Unlike the case of quench from the gapless phase, all the quasiparticle spectra of the pre-quench Hamiltonian satisfy $\varepsilon_{k}\cdot\varepsilon_{-k}\geq0$. Fig.~\ref{domain_DM} (b) and (c) show the domains $\mathcal{R}(h_{0},\gamma_{0})$ from the PM phase and FM$_{x}$ phase, respectively. For the quench from the point $(h_{0}=1.5,\gamma_{0}=0.2)$ in the PM phase, the domain ${\cal R}(1.5,0.2)$ covers the region $h<1$ [see Fig.~\ref{domain_DM} (b)], and leaves a small blank between $1<h<\sqrt{4D^{2}-\gamma^{2}+1}$. While for quench from the point $(h_{0}=0.5,\gamma_{0}=-0.5)$ in the FM$_{y}$ phase, a part of region in the gapless chiral phase region is not covered by the domain $\mathcal{R}(h_{0}=0.5,\gamma_{0}=-0.5)$ [see Fig.~\ref{domain_DM} (c)]. This reveals that the DQPTs might not occur in the case of quench the gapped phase to gapless chiral phase.

To explain why the DQPTs do not appear in the case of quench from the gapped phase to gapless phase, we display two examples of the Fisher zeros and the corresponding rate functions for the quench from the gapped phase in Fig.~\ref{DM_anisotropic_gapped}. The results in Figs.~\ref{DM_anisotropic_gapped} (a) and (b) correspond to the quench from $(h_{0}=1.5,\gamma_{0}=0.2)$ to $(h_{1}=1.02,\gamma_{1}=0.15)$ [see path E in Fig.~\ref{domain_DM} (b)]. While the results of Fisher zeros and rate function in Figs.~\ref{DM_anisotropic_gapped} (c) and (d) are obtained in the case of quench from $(h_{0}=0.5,\gamma_{0}=-0.5)$ to $(h_{1}=0.5,\gamma_{1}=-0.1)$ [see path F in Fig.~\ref{domain_DM} (c)]. It's found that the Fisher zeros of the two quench processes are located on the left of the complex time plane, and have no intersection with the imaginary axis [see Fig.~\ref{DM_anisotropic_gapped} (a) and (c)]. Unlike the cases of quenches from the gapless phase, here the Fisher zeros are not separated into two parts. The reason for the absence of DQPT is that the condition $|\tan{\alpha_{k}}|=1$ of (\ref{occur_condition}) is not satisfied in these quench processes, so that the lines of Fisher zeros do not cross the imaginary axis.

\section{the XY chain with three-site interaction}

As another example, we investigate the anisotropic \emph{XY} chain with the \emph{XZY-YZX} type of three-site interaction defined by the Hamiltonian \cite{Krokhmalskii2008,Liu_2012}
\begin{equation}\label{three-site}
 \begin{split}
    H & =\sum_{n=1}^{N}-\{\frac{1+\gamma}{2}\sigma^{x}_{n}\sigma^{x}_{n+1}+\frac{1-
    \gamma}{2}\sigma^{y}_
    {n}\sigma^{y}_{n+1}+h\sigma^{z}_{n} \\
      & +\frac{\xi}{4}(\sigma^{x}_{n-1}\sigma^{z}_{n}\sigma^{y}_{n+1}-\sigma^{y}_
      {n-1}\sigma^{z}_{n}\sigma^{x}_{n+1})\},
  \end{split}
\end{equation}
where  $\xi$ is the strength of the three-site interaction. By using the Jordan-Wigner transformation, the Hamiltonian (\ref{three-site}) can be written in the quadratic form (\ref{quardratic_H}) with
\begin{equation}\label{threesite_AB}
  \begin{array}{l}
    \begin{aligned}
      A_{mn}= & -h\delta_{mn}-\delta_{m,n+1}/2-\delta_{m+1,n}/2 \\
             &  -i\xi(\delta_{m,n+2}-\delta_{m+2,n})/4,
    \end{aligned}
     \\
    B_{mn}= \gamma\delta_{m,n+1}/2 - \gamma\delta_{m+1,n}/2.
  \end{array}
\end{equation}
The quasiparticle excitation spectra of the XY chain with three-site interaction are given by
\begin{equation}\label{Three_energy}
  \varepsilon_{k}=\xi\sin{2k}+2\sqrt{(\cos{k}+h)^{2}+\gamma^{2}\sin^{2}{k}}.
\end{equation}

\begin{figure}[t]
  \centering
  \includegraphics[width=1.0\linewidth]{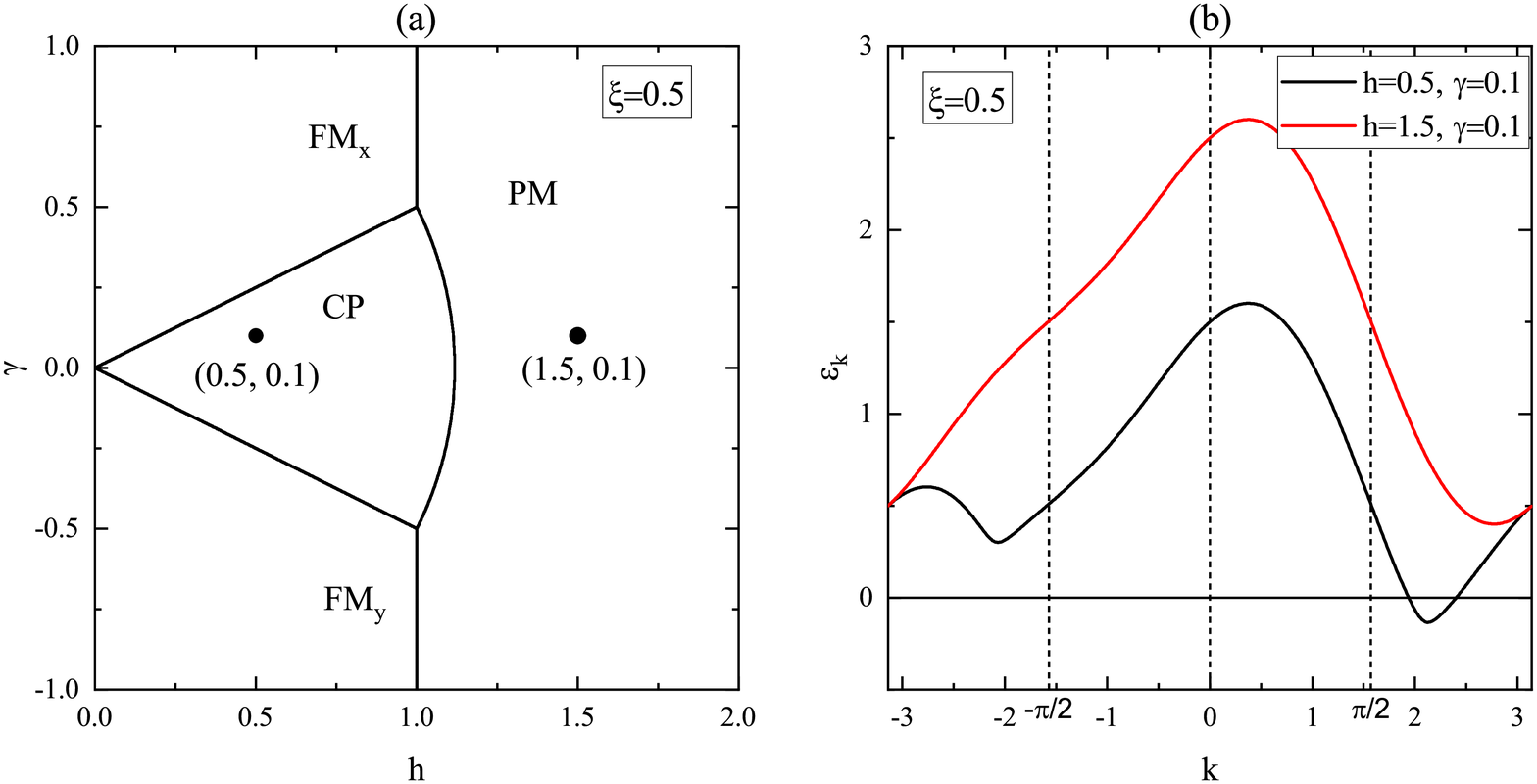}\\
  \caption{(a) The ground-state phase diagram of the system with $\xi=0.5$. Regions CP, PM, FM$_{x}$ and FM$_{y}$ between the solid lines correspond to the gapless chiral phase, gapped paramagnetic phase, gapped ferromagnetic phase along $x$ direction and gapped ferromagnetic phase along $y$ direction, respectively. (b) Typical quasiparticle excitation spectra of the Hamiltonian $H(0.5,0.1)$ and $H(1.5,0.1)$.  }\label{Three_phase}
\end{figure}

In Fig.~\ref{Three_phase} (a), we display a typical phase diagram of the system with three-site interaction $\xi=0.5$ \cite{Liu_2012}. The phase diagram contains four parts: the region CP is the gapless chiral phase, region PM is the gapped paramagnetic phase, region FM$_{x}$ is the gapped ferromagnetic phase along $x$ direction, FM$_{y}$ is the gapped ferromagnetic phase along $y$ direction, respectively. The critical lines separate the gapless phase and FM phase are given by $\gamma=\pm\xi h$, and that separates the gapless phase and PM phase is $h=\sqrt{\xi^{2}-\gamma^{2}+1}$. In Fig.~\ref{Three_phase} (b), we show two examples of quasiparticle excitation spectra, in which the Hamiltonian parameters $(h=0.5,\gamma=0.1)$ and $(h=1.5,\gamma=0.1)$ are marked by dots in Fig.~\ref{Three_phase} (a). The excitation spectra $\varepsilon_{k}$ are negative for $1.93<k<2.38$ when $h=0.5,\gamma=0.1$, while they are positive for all $k$ when $h=1.5,\gamma=0.1$.

\begin{figure}[t]
  \centering
  \includegraphics[width=1.0\linewidth]{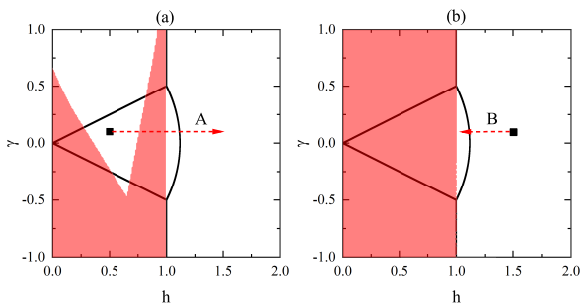}\\
  \caption{The domains ${\cal R}(h_{0},\gamma_{0})$ cover the post-quench Hamiltonian parameters where the DQPTs appear in the case of quench from the initial points $(h_{0},\gamma_{0})$. The initial points are marked by black squared dots: (a) $(h_{0}=0.5,\gamma_{0}=0.1)$, (b) $(h_{0}=1.5,\gamma_{0}=0.1)$,}\label{domain_three}
\end{figure}

Now, we consider the quantum quench from $H=H(h_{0},\gamma_{0})$ to $\tilde{H}=H(h_{1},\gamma_{1})$, where we keep the same strength of three-site interaction before and after the quench ($\xi_{0}=\xi_{1}$).

\begin{figure}[t]
  \centering
  \includegraphics[width=1.0\linewidth]{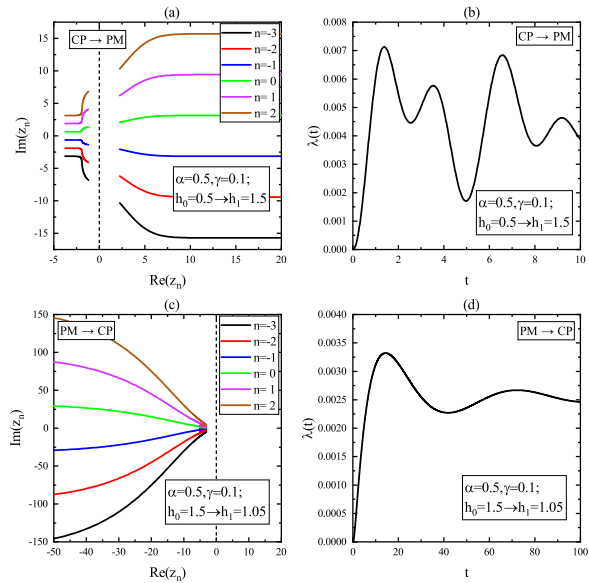}\\
  \caption{ (a) and (b) The Fisher zeros and corresponding rate functions for the quench from gapless chiral phase to the PM phase (CP$\rightarrow$PM), that is the path \textbf{A} seen in Fig.~\ref{domain_three} (a). (c) and (d) The Fisher zeros and corresponding rate function for the quench from the PM phase to the gapless chiral phase (PM$\rightarrow$CP), that is the path \textbf{B} seen in Fig.~\ref{domain_three} (b).}\label{Fisher2}
\end{figure}

\paragraph*{Quench from the gapless phase to gapped phase.} In Fig.~\ref{domain_three} (a), we show the domain ${\cal R}(0.5,0.1)$ which indicate that the DQPTs occur in the case of quench from the ground states of pre-quench Hamiltonian $H(0.5,0.1)$. It is found that the domain $\mathcal{R}(0.5,0.1)$ does not cover the area $h>1$, and leaves the blank blocks in the gapped FM$_{x}$ phase and FM$_{y}$ phase [see Fig.~\ref{domain_three} (a)]. This means that the DQPTs may not occur when the quench is from the ground state of $H(0.5,0.1)$ and across the QPTs. To make the behaviors of DQPTs more clearly, a typical example of Fisher zeros and rate function in the case of quench from the gapless phase to the gapped phase is shown in Fig.~\ref{Fisher2} (a) and (b). The corresponding quench path is from $(h_{0}=0.5,\gamma_{0}=0.1)$ to $(h_{1}=1.5,\gamma_{1}=0.1)$. We can see that the Fisher zeros $z_{n}(k)$ are separated into two branches and do not have interaction with the imaginary axis. The corresponding rate function is found to have no singularity point. The reason for the absence of DQPT is that the wave vectors corresponding to $\varepsilon_{k}\cdot\varepsilon_{-k}<0$ do not contribute to the Fisher zeros and rate functions.

\paragraph*{Quench from the gapped phase to gapless phase.} We also consider the case of quench from the gapped phase. In Fig.~\ref{domain_three} (b), we display the domain ${\cal R}(1.5,0.1)$ for the occurrence of DQPTs in the case of quench from the ground state of pre-quench Hamiltonian $H(1.5,0.1)$ which is in the gapped PM phase. It's found that the domain ${\cal R}(1.5,0.1)$ only covers the region $h\leq1$ and leaves a small blank area $1<h<h_{c}=\sqrt{\xi^{2}-\gamma^{2}+1}$ in the gapless phase. This indicates that there is a possibility that DQPTs do not appear in the case of quench from the gapped phase to gapless phase. Similarly, we show an example of Fisher zeros and rate function in the quench from gapped phase to gapless phase in Fig.~\ref{Fisher2} (c) and (d). The corresponding quench path is from $(h_{0}=1.5,\gamma_{0}=0.1)$ to $(h_{1}=1.05,\gamma_{1}=0.1)$. It can be seen that unlike that in the case of quench from the gapless phase, the Fisher zeros $z_{n}$ coalesce to continuous line and do not be cut into two branches, because all excitation spectra of pre-quench Hamiltonian in the gapped phase satisfy $\varepsilon_{k}\cdot\varepsilon_{-k}>0$ and contribute to the Fisher zeros. The reason for the absence of DQPTs is that the condition $|\tan{\alpha_{k}}|=1$ of (\ref{occur_condition}) are not satisfied.

\section{Summary and conclusion}

We study the properties of DQPTs in general quantum spin chains, in which the systems have gapless phases and asymmetrical quasiparticle excitation spectra. By considering the quench starting from different initial states, we find that the factors $\mathcal{L}_{k}(t)$ of LE equal unity where the quasiparticle excitation spectra of pre-quench Hamiltonian satisfy $\varepsilon_{k}\cdot\varepsilon_{-k}<0$. Therefore, we obtain the general conditions for the occurrence of DQPT, that is the occurrence of DQPT not only requires the Bogoliubov angle to satisfy $|\tan{\alpha_{k}}|=1$, but also requires the quasiparticle excitation spectra of the pre-quench Hamiltonian to satisfy $\varepsilon_{k}\cdot\varepsilon_{-k}\geq0$. We have confirmed our conclusions in two typical models---the \emph{XY} chains with DM interaction and \emph{XZY-YZX} type of three-site interaction.

From the discussion above and previous results \cite{Jafari2019,Haldar2020}, we summarize the connection between DQPT and QPT as following. For the system only with gapped phase, the DQPTs will occur when the quench crosses the critical lines of QPTs. For the system with gapless phase and symmetrical quasiparticle excitation spectra, the DQPTs may not occur in the quench from the gapped phase to gapless phase but occur in the quench from the gapless phase to gapped phase. For the system with gapless phase and asymmetrical quasiparticle excitation spectra,  the DQPTs may not occur in the quench across the quantum phase transitions regardless of whether the quench is from the gapless phase to gapped phase or from the gapped phase to gapless phase.

\begin{acknowledgments}
  The  work was supported by the National Natural Science Foundation of China (Grant Nos. 11975126 and 11575087).
\end{acknowledgments}

\bibliography{dqpt_asymmetry}

\end{document}